\documentclass[12pt,a4paper]{article}
\usepackage[T2A]{fontenc}
\usepackage[cp1251]{inputenc}
\usepackage[english]{babel}
\usepackage{hyperref,cite,a4wide}

\usepackage{amsmath,amssymb,amsthm,amsfonts,amscd}

 \theoremstyle{plain}
\newtheorem{theorem}{Theorem}
\newtheorem*{main*}{Main Theorem}
\newtheorem{lemma}[theorem]{Lemma}

\newtheorem{corollary}[theorem]{Corollary}

\theoremstyle{definition}

\newtheorem*{remark*}{Remark}
\newtheorem*{example*}{Example}

\def\R{{\Bbb R}}

\def\A{{\bf A}}

\def\W{{\bf C}}

\def\x{{\bf x}}

\def\b{{\bf b}}

\def\grad{{\bf{grad}}}
\def\div{{\operatorname{div}}}
\def\vol{\bf{vol}}
\def\rot{{\operatorname{curl}}}
\def\vol{{\bf{vol}}}

\def\R{{\Bbb R}}
\def\B{{\bf B}}
\def\b{{\bf b}}
\def\A{{\bf A}}

\begin{document}

\title{\textbf{Evolution of Mirror Axion Solitons}}

\author{P.~M.~Akhmetiev\thanks{pmakhmet@mail.ru}, M.~S.~Dvornikov\thanks{maxim.dvornikov@gmail.com}
\\
\small{\ Pushkov Institute of Terrestrial Magnetism, Ionosphere} \\
\small{and Radiowave Propagation (IZMIRAN),} \\
\small{108840 Moscow, Troitsk, Russia}}

\maketitle

\begin{abstract}
We study an axion soliton, which weakly interacts with background matter and magnetic fields. A mirror-symmetric soliton, for which the magnetic flow is due to secondary magnetic helicity invariant, is described by the Iroshnikov-Kreichnan spectrum.
For a large scale magnetic field dynamo is not observed. In a mirror axionic soliton, a phase transition, which produces a magnetic helical flow,
is possible. Using this transition, the soliton becomes mirror-asymmetric. When the mirror symmetry is broken, the axion soliton allows the magnetic energy, which is the result of the transformation of the axionic energy. In the main result, for an initial stage of the process, we calculate a scale for which the generation of large scale magnetic fields is the most intense. By making numerical simulations, we received that lower lateral harmonics of the magnetic field have greater amplitudes compared to higher ones.  A simplest statistical ensemble, which is defined by the projection of all harmonics onto principal harmonics is constructed. We put forward an assumption that it was the indication to some instability in axionic MHD. Now, we can provide a possible explanation of this feature. When the mirror symmetry of the axion soliton is broken, the $\gamma$-term in the
axionic mean field equation, which is related to the axion spatial inhomogeneity, interacts with principal harmonics. As the result, the axion soliton acquires the magnetic energy and becomes helical.
\end{abstract}


\section{Introduction}

According to modern astronomical observations, the major fraction of the gravitating mass in the universe is present in the form of dark matter. Masses of dark matter constituents can vary quite significantly (see, e.g., Ref.~\cite{Tis07}). Despite the issue of the dark matter composition is still open, it is believed to consist of rather light particles with masses $m \lesssim 10^{-4}\,\text{eV}$, which are called axions and/or axion like particles (ALPs). Originally, axions were proposed in Refs.~\cite{PecQui77,Wei78,Wil78} within the quantum chromodynamics (QCD) to address the strong CP problem which, in particular, would lead to a rather great electric dipole moment of a neutron that has not been observed yet; cf. Ref.~\cite{Abe20}.

Later, it was understood, e.g., in Ref.~\cite{DinFis83}, that axions and ALPs can be created in sizable amounts to explain the dark matter contribution to the total mass of the universe. Under some conditions, an axion background can be spatially inhomogeneous leading to the formation of axion clusters~\cite{KolTka93,KolTka94}
and axion stars~\cite{BraZha19,Vis21}. The collisions of axion stars with other astrophysical objects are reported in Ref.~\cite{BraZha19} to result in various multimessenger astronomy effects.

Despite an axion does not have an electric charge, such a particle can quite weakly interact with electromagnetic fields (see, e.g., Ref.~\cite{KimCar10}). Thus, we can consider the axion electrodynamics. If one deals with large scale magnetic fields in a conducting medium in presence of axions, one can refer to the axion magnetohydrodynamics (MHD). Magnetic fields under the influence of an axion background were found in Ref.~\cite{LonVac15} to be unstable. We studied the axion MHD in the early universe in Refs.~\cite{DvoSem20,Dvo22}. The axionic MHD in neutron stars was considered in Ref.~\cite{Anz23}. The instability in a laser beam interacting with axions was studied in Ref.~\cite{Bey23}.

Recently, in Refs.~\cite{D-A,AkhDvo24}, we derived the modified induction equation for magnetic fields interacting with spatially inhomogeneous axions. Based on this new equation, in Refs.~\cite{Dvo24,Dvo25}, we described the evolution of magnetic fields in a dense axion clump embedded in solar plasma. This behavior of magnetic fields was suggested in Ref.~\cite{Dvo24} to have the implication to the solar corona heating problem. Indeed, the temperature of the solar corona is about three orders of magnitude higher than that of the solar surface. Classical electrodynamics faces certain difficulties in the explanation of the observed corona temperature (see, e.g., Ref.~\cite{Kli15}), i.e. it is an open issue in solar physics.

The alternative mechanism for the solar corona heating, involving dark matter particles, was suggested in Ref.~\cite{DilZio03}. Then, this idea was further developed in the series of works by Zhitnitsky (see, e.g., Ref.~\cite{Zhi17}), who suggested that the solar corona is heated up by decaying axion quark nuggets. This structures were assumed to be stored in solar interiors and float up to the solar surface.

The mechanism for the corona heating, proposed in Refs.~\cite{Dvo24,Dvo25}, also involves axionic clumps. Nevertheless, it is different from that in Ref.~\cite{Zhi17} since one deals with magnetized axion structures in Refs.~\cite{Dvo24,Dvo25}. The important problem left unexplored in Refs.~\cite{Dvo24,Dvo25} is the dependence of the magnetic field generation on the length scale of the system. This issue is addressed in the present work. Moreover, developing an alternative model of the solar corona heating can constrain the characteristics of axion quark nuggets which are assumed to be a quite exotic form of dark matter.

In the present work, we continue our studies of the magnetic fields evolution driven by inhomogeneous axions by considering the asymptotic ergodic invariants of magnetic lines.
In Ref.~\cite{D-A1}, such an analysis allowed one to investigate a contribution of the curvature parameter of the expended 3D infinite space on the  $\alpha$-effect, 
which is observed for magnetic field after the electroweak phase transition in Early Universe, when
elementary particles acquire masses.  
We assume that, during an unknown phase transition, the Chern-Simons invariant of solutions of
Yang-Mills equations has to preserved and is transformed into the $M$-invariant of a magnetic flow
of a large scale magnetic field, which is a Chern-Simons field. 
Such an invariant has a density, which is preserved by an ideal small-scale transformation
of the magnetic field in a liquid conductive domain. 
 The curvature of the expanding space is related with the invariant of a large scale magnetic field. In its turn, the invariant of magnetic lines is related to the Kolmogorov low, which characterizes the MHD system. It allows us to conclude the existence of the $\alpha$-effect
  without exact solutions during the
 phase translation, as well as without the effect of the gravity on the curvature.
 
We apply an analogous idea to study axion solitons.
Namely, we consider a process of the transformation of a mirror axion soliton into helical one. The mirror axion soliton is defined using a special configuration of magnetic field in the form of the Hopf fibration, which was constructed in
Ref.~\cite{D-A}. A magnetic flow of the axion soliton is defined by the contribution of the first term in the mean-field equation; see Eq.~(\ref{meaN}) below. By this hypotesis, one may conclude that 
a  total spectrum of the soliton, which includes the contribution of energy of axion field,
has to coincide with the Iroshnikov-Kreichnan spectrum with the power
 $\tau \sim -\frac{3}{2}$. This is the only spectrum which magnetic flow is characterized by  quadratic magnetic helicity (see Refs.~\cite{A1,D,S}). It means that, in this flow, left- and right-linked magnetic lines  have equal contributions.  

The magnetic energy of a mirror axion soliton could be arbitrary small. The total energy is defined by the energy of axion field, that is not small. The evolution of mirror axion soliton
 is characterized by the invariant density of the quadratic magnetic helicity.
 
By a bifurcation of a mirror axion soliton into a polarized soliton, the Iroshnikov-Krechnan spectrum is transformed into the Kolmogorov spectrum. The magnetic flow becomes (left- or right-)polarized. By the Arnol'd inequality~\cite{Arn} (see Eq.~(\ref{BB}) below), 
 the magnetic energy flow is bounded from below.  
 The power of the spectrum is transformed in the following way: 
$$\tau^{fl}=-1 \mapsto \tau^{fl}=-\frac{7}{6}. $$
Thus, the magnetic helicity flow generates the magnetic energy flow.

The Arnold inequality gives a lower bound of $L^2$-norm of magnetic field,
which is equal to the lower bound of the magnetic energy, by the absolute value of the magnetic helicity.
We prove a generalization of this inequality, in particular, for mirror magnetic configurations. 
The generalized Arnold inequality relates  $L^4$-norm of magnetic field with the quadratic magnetic helicity. In mirror configurations, lower bonds of magnetic energy do not exist. Thus,
the magnetic energy could be an arbitrary small in the case when magnetic lines are sufficiently thin. The absolute value of the magnetic helicity is also small since,
for linking coefficients of magnetic lines, which are proportional to the product of magnetic flows
trough lines, the magnetic energy could be arbitrary small.

The Arnold inequality expresses one of the main basic idea in dynamo theory about a balance
of the magnetic energy and the magnetic helicity. This key inequality is not  mentioned
 often since analogous idea was formulated by different authors in different ways.  
Our approach towards the Arnold inequality is new. It is different from the known approach based on Fourier series.
A generalized Arnold inequality was investigated in 
 Ref.~\cite{A-K-S}  (see also references therein) based on the same idea with Fourier harmonics. However, in the required form, it was not proved anywhere.  

To calculate the quadratic magnetic helicity for magnetic field in a compact domain, 
we have to take the sum of squares of pairwise linking coefficients. By this construction, the contributions of a pair of magnetic lines 
are taken with the square of magnetic flows. In this case, magnetic energy could be small. However, the $L^4$-norm of magnetic field is bounded from below by the quadratic magnetic energy and cannot be small if quadratic magnetic energy is sufficiently large.
  
The phase transition of a mirror axion soliton into polarized axion soliton is complicated and could be investigated using more elaborated techniques compared to the mean field theory. 
We provide examples where means-values operators of large scale fields from small-scale fields are not commuted with the vorticity operator of small-scale fields. 
In addition to a simplest example from Sec.~\ref{sec2}, when this effect is related with time-mean values, in Appendices~\ref{sec5} and~\ref{sec6} we give additional two  examples, which just demonstrate that the phase transition of axion solitons is complicated.
The first example is related with Maxwell multipole theorem
 in Ref.~\cite{Arn2}. By this construction, Green functions of small-scale magnetic dipoles are defined. The second example the effect is related with the idea of Mischenko and Fomenko for a local formula. More precisely, for the formula of shifts, which is discovered in 
  Ref.~\cite{M-F}, which is applied for invariant density of magnetic helicity~\cite{A2}.
  There are additional difficulties since the Mischenko-Fomenko formula is applied for polynomial. However, the magnetic helicity density is defined by an infinite series. 
  An interesting fact that the mean value of the infinite raw is not equal to the mean value of the principal term of the series. It gives one no divergence of the mean value. However,
 it gives a correction of the mean value of the order
  $\sim \mathcal{O}(1)$.

The present work is organized in the following way. In Sec.~\ref{sec2}, we give simplest arguments, which demonstrate that the concept of magnetic lines of a large scale magnetic field for phase-transitions is complicated. In Sec.~\ref{sec3}, we recall the Arnold inequality~\cite{Arn} and prove its generalization. The main result is obtained in Sec.~\ref{sec4}. This result is related to the question of the evolution of an axion soliton.
 
\section{Mean large scale magnetic flows of small-scale magnetic field could be differ from a flow of large scale field}\label{sec2} 

To work with a large scale magnetic field in two-scale dynamo theory, a concept of statistic ensembles is developed in Ref.~\cite{Mo}. Let us consider one example related to the problem. 

Assume that, in a small-scale ensemble of an axion solitons (see Secs.~4 and~5 in
 Ref.~\cite{D-A} for the description of the axion soliton), one has the magnetic modes
 $\B_A$ and $\B_B$. We also assume that the mode $\B_A$ does not depend not on time (or depends on time with the frequency $\omega_A=0$), whereas the mode $\B_B$ oscillates in time with the non-zero
 frequency  $\omega_B$. Assume that the given modes are produced as mean modes in a simple statistic ensemble $\Omega_1$ (of principal harmonics with no lateral coordinate) of corresponding modes by the convolution of solutions 
 of the total ensemble $\Omega_2 \mapsto \Omega_1$  (an explicit description of $\Omega_2$ is unknown). 
 The ensemble $\Omega_1$ is constructed by the spacial coordinate 
  $\theta$ and the time $t$. 
  
  Let us denote by $\Omega$ the ensemble with the only spacial coordinate  $\theta$, the
  mean operator over the time is denoted by  $\Omega_1 \mapsto \Omega$. For times  $T > \omega_B^{-1}$ in the ensemble  $\Omega$ the mode  $\B_A$ is observed, but the mode  $\B_B$ is absent. The mean value of the magnetic helicity  $\chi_{\Omega_1}$ equals to zero, but the quadratic magnetic helicity  (see Refs.~\cite{A1,S,D}) admits an appropriate value
   $\chi^{[2]}_{\Omega_1}$.

Assume that, in axion ensemble 
 $\Omega_1$, we get an axion  $\varpi(\theta,t)$ with the same frequency  $\omega_{\varphi}$. Let us change the parameter $\omega_{\varphi}$ on the segment $[0,\omega_B]$, accordingly with the Eq.~(5.3) in Ref.~\cite{D-A}. Then, for
 $\omega_{\varphi}=0$, the magnetic helicity flow in the ensemble  $\Omega_1$,
 caused by the first term in Eq.~(2.5) in Ref.~\cite{D-A} and calculated by Eq.~(5.5) in Ref.~\cite{D-A} becomes zero.  But in the case $\omega_{\varphi} = \omega_B$ the magnetic helicity flow in the ensemble $\Omega_1$ is non-trivial since two linked mode exist. 
 The quadratic helicity flow in the ensemble
  $\Omega_1$ depends on of the value of the parameter  $\omega_{\varphi}$.

This example shows that by variations of internal parameters of a complicated axion soliton
mirror symmetry can be distroyed and the magnetic helicity flow can be observed. 
In the evolution equation in the ensemble
 $\Omega$, the given change is not observed since the averaging operator  $\Omega_1 \mapsto \Omega$ does not depend on the internal parameters. 

\section{ The Arnold inequalities in the case of bounded domains}\label{sec3}

The Arnold inequality estimates from above the magnetic energy by the absolute value of the helicity integral
$\chi_{\B}=\int_{\Omega} (\A(\x),\B(\x)) dV$. The magnetic energy (versus the full energy) is not an invariant
for the ideal MHD, but this lower bound is given in Ref.~\cite[Ch.~III, Theorem~1.5]{A-Kh}.

\begin{theorem}\label{Arnold}
	Let $\B(\x)$ be a magnetic field in a domain  $\Omega$, which is tangent to the boundary surface in the case
	$\partial(\Omega) \ne \emptyset$. The following inequality is satisfied: 
	\begin{eqnarray}\label{BB}
		U(\B) = \int_{\Omega} (\B(\x),\B(\x)) dV \ge C \left\vert \int_{\Omega} (\A(\x),\B(\x)) dV \right\vert , 
	\end{eqnarray}
	where 
	$C$ is a positive constant of the dimension $\text{cm}^{-1}$, which depends only on a size  and of a form of the domain $\Omega$. 
	For a domain of the diameter $d$, one can put $C = d^{-1}$. In a special case, when $\Omega$ is a ball of the radius $R$
	we may prove a more strong inequality with $C > d$ by an explicit calculation. When $\Omega$ is a thin ellipsoid, which is concentrated near the big axis (a one-parameter family of ellipsoids tends to the segment) we get $C \to 0+$.
	
\end{theorem} 

The multiple integral in the right side of the formula 
$(\ref{BB})$ is called the magnetic energy.

\subsubsection*{Proof 1 of Theorem \ref{Arnold}}
The Arnold inequality is a corollary of the Cauchy-Bunyakovsky inequality and the Poincar`e inequality.

Let us apply the Cauchy-Bunyakovsky inequality:
$$ \left(\int_{\Omega} (\B(\x),\A(\x)) dV \right)^2 \le \int_{\Omega} (\A(\x),\A(\x))dV \int_{\Omega} (\B(\x),\B(\x)) dV. $$
Then, from the Poincar\`e inequality, we get:
$$ \int_{\Omega} (\A(\x),\A(\x)) dV \le \frac{1}{C^2} \int_{\Omega}(\B(\x),\B(\x)) dV. $$

The Poincar\'e inequality is proved by the Fourier series expansion of
$\B(\x)$ over eigenvector functions of the vorticity operator $\rot$ in a bounded domain  $\Omega$. The constant $C$ is the smallest absolute proper value and has the dimension cm$^{-1}$.  In the case of periodic magnetic field in $\R^3$ the constant 
$(2\pi)C^{-1}$  is the greater period of the vector-function $\rot$. \qed  

For a bounded domain $\Omega \subset \R^3$ the constant $C$ is complicated. In particular, in the case $\Omega \subset \R^3$ is a ball of a prescribed radius, the spectrum of the operator $\rot$ is knows. But in a general case this problem is not solved. We present an alternative proof
of the Arnold inequality.

\subsubsection*{Proof 2 of Theorem \ref{Arnold}}
Let us used the Biot-Savart low, which express the vector
$\A(\x)$ 
by the vector $\B(\x_1)$:
$$\A(\x) =  \int_{\Omega} \frac{1}{4\pi} \frac{\B(\x_1) \times (\x_1 - \x)}{\vert \x_1 - \x \vert ^3} d\x_1.$$

The Biot-Savart low  is  the following formula:
$$\A(\x) = \int_{\Omega} \A(\x;\x_1) d\x_1,$$
with denotations:
$$ \A(\x;\x_1)= \frac{1}{4\pi} \frac{\B(\x_1) \times (\x_1 - \x)}{\vert \x_1 - \x \vert ^3}. $$

By the Biot-Savart formula we get:
$$ \left(\A(\x),\A(\x)\right) = \left(\int_{\Omega} \A(\x;\x_1) d\x_1,\int_{\Omega} \A(\x;\x_2) d\x_2\right) \le $$
$$(4\pi)^{-2}\left(\int_{\Omega_1} \vert \x - \x_1 \vert^{-2}\vert\B(\x_1)\vert d\x_1,\int_{\Omega_2} \vert \x - \x_2 \vert^{-2}\vert\B(\x_2)\vert d\x_2\right) = $$
$$(4\pi)^{-2}\iint_{\Omega_1 \times \Omega_2}\vert \x - \x_1 \vert^{-2} \vert \x - \x_2 \vert^{-2} \vert \B(\x_1)  \vert  \vert \B(\x_2)  \vert d \x_1 d \x_2 \le$$
$$(4\pi)^{-2}\iint_{\Omega_1 \times \Omega_2}\vert \x - \x_1 \vert^{-2} \vert \x - \x_2 \vert^{-2}\B^2(\x_2)d\x_1 d \x_2 =$$
$$(4\pi)^{-2}\int_{\Omega_1}\vert \x - \x_1 \vert^{-2} d \x_1 \int_{\Omega_2} \vert \x - \x_2 \vert^{-2} \B^2(\x_2) d \x_2.$$
We use the inequality: $\vert  \B(\x_1)  \vert  \vert \B(\x_2) \vert \le \frac{1}{2}(\B^2(\x_1)+\B^2(\x_2))$. In the formula above, $\Omega_1$ and
$\Omega_2$ are the two copies of the domain $\Omega$.

Because $\int_{\Omega_1}\vert \x - \x_1 \vert^{-2} d \x_1 \le 4\pi d$, where $d$ is the diameter of the
domain $\Omega$, we get:
$$\int_{\Omega_1}\vert \x - \x_1 \vert^{-2} d \x_1 \int_{\Omega_2} \vert \x - \x_2 \vert^{-2} \B^2(\x_2) d \x_2 \le$$
$$4\pi d \int_{\Omega_2} \vert \x - \x_2 \vert^{-2} \B^2(\x_2) d \x_2. $$
$$ \int_{\Omega} (\A(\x),\A(\x)) d\x \le d \int_{\Omega \times \Omega_2} \vert \x - \x_2 \vert^{-2} \B^2(\x_2) d \x d \x_2 \le$$
$$ d^2 \int_{\Omega_2} \B^2(\x_2) d\x_2 = \frac{1}{C^2} \int_{\Omega}(\B(\x),\B(\x)) dV, $$
where $C = d^{-1}$. 

In the case $\Omega$ is a ball of the radius $R$, we get:
$\int_{\Omega_1}\vert \x - \x_1 \vert^{-2} d \x_1 < 4\pi d$. In the case $\Omega \subset U_{\varepsilon}[0,1]$, we get:
$\int_{\Omega_1}\vert \x - \x_1 \vert^{-2} d \x_1 \approx \varepsilon.$
\qed

Using denotations $U_{\B}=\int_{\Omega}(\B(\x),\B(\x))d\x$ for magnetic energy and
$\chi_{\B} = \int_{\Omega} (\A(\x),\B(\x)) dV$ for the magnetic helicity integral, we can rewrite the inequality in Eq.~(\ref{BB}) as follows:
$$ U_{\B} \ge C \vert \chi_{\B} \vert . $$

Using Eq.~(\ref{BB}), we can write down the following inequality for the Holder $4$-norm
of the magnetic energy: $U^{(4)}(\B)=\int_{\Omega} (\B(\x),\B(\x))^2 d\x$. By the  Cauchy-Bunyakovsky inequality,
we get: 
$$ vol(\Omega) \int_{\Omega} (\B(\x),\B(\x))^2 d\x \ge \left(\int_{\Omega} (\B(\x),\B(\x)) d\x \right)^2. $$
Then by the Arnold inequality, we get:
\begin{eqnarray}\label{Arnold3}
	vol(\Omega) U^{(4)}_{\B} \ge C^2 \chi_{\B}^2. 
\end{eqnarray} 
This inequality works only for polarized magnetic fields. In the cases $\chi_{\B} \approx 0$ of magnetic fields, which are not left-right polarized, the left-hand part of the inequality equals zero and the inequality becomes trivial.
Let us prove a more strong inequality for $U^{(4)}$, which works even in the case $\chi_{\B}=0$ when the magnetic helicity equals zero.

\begin{theorem}\label{Arnold2}
	
	Let $\B(\x)$ be a magnetic field in a domain  $\Omega$, which is satisfies the boundary conditions from Theorem \ref{Arnold}. The following inequality is satisfied:
	\begin{eqnarray}\label{BB}
		U^{(4)}_{\B} = \int_{\Omega} (\B(\x),\B(\x))^2 dV \ge C^2 \int_{\Omega} (\A(\x),\B(\x))^2 dV, 
	\end{eqnarray}
	where 
	$C$ is a positive constant of the dimension $cm^{-1}$ as in Theorem \ref{Arnold}.
\end{theorem} 

\begin{corollary}\label{Corr}
	The inequality in Eq.~(\ref{Arnold3}) results from Eq.~(\ref{BB}).
\end{corollary}

\subsection*{Proof of Corollary \ref{Corr}}
The function $(\A(\x),\B(\x))^2$, where $\x \in \Omega$, is called the correlation tensor for the quadratic helicity density.
We have the inequality 
\begin{eqnarray}\label{Uchi(2)}
	\int_{\Omega} (\A(\x),\B(\x))^2 d\x \ge \int_{\Omega} m[(\A(\x),\B(\x))^2] d\x = \chi^{(2)}_{\B},
\end{eqnarray}
where the quadratic magnetic helicity density $m^2[(\A(\x),\B(\x))]$ is the ergodic mean value of $(\A(\x),\B(\x))^2$
with respect to the magnetic flow, $\chi^{(2)}_{\B}$ is the quadratic magnetic helicity integral. 
Therefore we get:

Then, using the inequality
\begin{eqnarray}\label{chi(2)}
	\chi^{(2)} \ge  \vol^{-1}(\Omega) \chi_{\B}^2, 
\end{eqnarray}
between the quadratic magnetic helicity  and the magnetic helicity, we get Eq.~\eqref{Arnold3}. \qed

\subsection*{Proof of Theorem \ref{Arnold2}}
By the Biot-Savart formula we get:
$$ (\A(\x),\A(\x))^2  \le $$
$$(4\pi)^{-4}\int_{\Omega_1}\vert \x - \x_1 \vert^{-2}d\x_1\int_{\Omega_2} \vert \x - \x_2 \vert^{-2}d \x_2$$
$$\times\int_{\Omega_3}\vert \x - \x_3 \vert^{-2}d\x_3\int_{\Omega_4} \vert \x - \x_4 \vert^{-2}\B^4(\x_4)d \x_4$$
We use the inequality: $\vert \B(\x_1)  \vert \vert\B(\x_2)\vert \vert\B(\x_3) \vert \vert \B(\x_4)\vert  \le \frac{1}{4}(\B^4(\x_1)+\B^4(\x_2)+\B^4(\x_3)+\B^4(\x_4))$.

Therefore
$$ \int_{\Omega}(\A(\x),\A(\x))^2d\x \le  C^4 \int_{\Omega_4} \B^4(\x_4) d\x_4 , $$
where 
$$ C^4 =  (4\pi)^{-4}\int_{\Omega_1}\vert \x - \x_1 \vert^{-2}d\x_1\int_{\Omega_2}\vert \x - \x_2 \vert^{-2}d\x_2\int_{\Omega_3}\vert \x - \x_3 \vert^{-2}d\x_3\int_{\Omega}\vert \x - \x_4 \vert^{-2}d\x.$$
The constant $C$ has the dimension cm$^{-1}$, we can put $C = d$ as in Theorem~\ref{Arnold}. \qed.

\section{Mean-field equation with the axionic $\gamma$-term}\label{sec4}

In this section, we investigate an example in which the time average operator the magnetic helicity flow is vanishing. However, the flow of quadratic magnetic helicity exists.
Since the quadratic helicity density in the ideal approximation is an invariant function, which is frozen in to a liquid medium, this value tends to a constant in the domain and
determines a power of the spectrum for the simplest axion statistic ensemble $\Omega$.
This power coincides with the power of the Eroshnikov-Kreichnan spectrum, which is explained in Ref.~\cite{Z-R-S}. 

The mirror soliton, for which magnetic dynamo is not possible, is determined by the  
quadratic helicity density. As the result of a phase transition, this soliton can produce a magnetic helicity flow, which is, from the physical point of view, is presented by the Kolmogorov spectrum, which determines the most natural energy spacial distribution. 
It is remarkable that the Kolmogorov spectrum determines the uniformly large scale  density distribution
of the $M$-invariant of magnetic lines, which is invariant for small scale magnetic lines in the ideal approximation. For the Kolmogorov spectrum, the axion energy is partially transformed into the magnetic energy since, by the Arnold inequality, a helical magnetic helicity flow have to be equipped with a flow of the magnetic energy.

Let us present the following calculation of the wave number for the mean-field equation with the axionic $\gamma$-term, for which the magnetic flow caused by the $\gamma$-term is the largest. 

In Refs.~\cite{D-A,AkhDvo24}, we have derived the new induction equation accounting for the axions contribution,
\begin{equation}\label{meaN}
  \frac{\partial \B}{\partial t} = \rot (\b \times \rot \B) +\alpha \rot \B - \eta \rot \rot \B,
\end{equation}
where $\alpha = g_{a\gamma} \eta \partial_t \varphi$ is the $\alpha$-dynamo parameter, $\mathbf{b} = g_{a\gamma} \eta^2 \nabla \varphi$ is the new axial vector term which is nonzero for inhomogeneous axions, $\varphi = \varphi(\mathbf{x},t)$ is the axion wave-function, $g_{a\gamma}$ is the axion-to-photon coupling constant, and $\eta$ is the magnetic diffusion coefficient. We mention the difference between Eq.~\eqref{meaN} and the induction equation used in Ref.~\cite{DvoSem20}, where the axion field was assumed to be spatially homogeneous. Therefore, the first term in the right hand side, $\rot (\b \times \rot \B)$, was absent in Ref.~\cite{DvoSem20}.  We call this new contribution to the induction equation as the $\gamma$-term, because the second and the third terms are called $\alpha$ and $\beta$ terms correspondingly.

We assume that, in Eq.~(\ref{meaN}), only mean values are kept, which are characterized by the $3D$-inhomogeneous distribution
of the axion field. Since we apply the mean field equation for the simplest statistic ensemble 
$\Omega$, we have to assume that the values  $\alpha$ and $\eta$ in this equation depend only on the scale. We found in Refs.~\cite{Dvo24,Dvo25} by means of numerical simulations that the typical period of axions oscillations is much larger than the time scale of magnetic fields variation. That is why we can neglect $\alpha$ in Eq.~\eqref{meaN}. 
We assume that the effect is due to the 
 axionic $\gamma$-term and due to turbulent diffusion $\beta$-term in Eq.~\eqref{meaN}, $\propto \eta \rot \rot \B$. This term depends on the scale by the Kolmogorov's law.

Let us pass to dimensionless terms in Eq.~(\ref{meaN}) as in Eq.~(2.4) in Ref.~\cite{Dvo25}. For simplicity, we will not introduce new notations for normalized values. For the Kolmogorov spectrum, the amplitude of the magnetic vector
depends on dimensionless wave number as 
 $\B \propto k^{-\frac{5}{6}}$. Therefore, by the magnetic and axionic energy balance, we get that  $\varphi \propto k^{-\frac{5}{6}}$. 

Instead of one dimensional distributions, now, we consider convolutions of observables having power corresponding to the $3D$-distribution of vectors in Eq.~(\ref{meaN}). For the Kolmogorov spectrum, the amplitude of the magnetic vector depends on the dimensionless wave number, and because
 $\b \propto \nabla{\varphi}$, we get:
\begin{eqnarray}\label{b}
	\b \propto k^{-\frac{11}{6}}. 
\end{eqnarray}
For the turbulent diffusion, we have to take
$\propto \eta k^{-\frac{1}{3}}$ owing to the Kolmogorov law-$\frac{2}{3}$ for a mean distribution of
the square of the hydrodinamic velocity. Since $\eta$ is a dimensionless turbulent magnetic diffusion coefficient, which is characterized by a distribution of a large scale magnetic field from its ideal approximation. Let us substitute this asymptotic in the mean field equation.
For 
the $\beta$-term, one has that $\B^{flow;\beta} \propto k^{-\frac{7}{6}}$, as it was shown in Ref.~\cite{D-A1}. 
For the $\gamma$-term, accordingly with the contribution of the axionic field instead of turbulent diffusion term, one gets that $\B^{flow;\gamma} \propto  k^{-\frac{7-2+11}{6}} = k^{-\frac{8}{3}}$.

The magnetic flow versus the wave number
 $k$ reads
$$ \propto k^{-\frac{8}{3}}\eta^{-1} - k^{-\frac{7}{6}}, $$
which is opimized for positive $k$, in the case when $\eta$ is sufficiently small. We get the expression,
$$ k_{max} = \left( \frac{16}{7\eta} \right)^{\frac{2}{3}} \propto \eta^{-\frac{2}{3}}. $$

The magnetic flow and the magnetic field for an initial system, which is described by
the Iroshnikov-Kreichnan spectrum, are small, because the energy is concentrated in the axionioc component. The Arnold inequality gives a restriction from below of $L^4$-norm of magnetic field. However, it gives no restriction from below of the magnetic energy.

After the mirror symmetry of the axion soliton is broken, the magnetic flow is concentrated at a prescribed scale, if the coefficient of turbulent diffusion $\eta$ is sufficiently small. The axion soliton itself acquires the magnetic energy.

\section{Conclusion}

We have investigated the axion soliton, which weakly interacts with background matter and magnetic fields, using 
higher invariants of magnetic lines. A mirror-symmetric soliton, for which the magnetic flow
is due to magnetic helicity invariant, is described by the Iroshnikov-Kreichnan spectrum.
For a large scale magnetic field dynamo is not observed.

In a mirror axionic soliton, a phase transition, which produces a magnetic helical flow,
is possible. Using this transition, the soliton becomes mirror-asymmetric.
When the mirror symmetry is broken, the axion soliton allows the magnetic energy, which is the result of the transformation of the axionic energy.

A future detailed consideration is possible using the invariants of magnetic lines, such as the $M$-invariant. Our study describes the evolution of the system.
We use the quadratic magnetic helicity, which is a second-moment of the magnetic helicity that can be observed in mirror systems. 
In the main result in Sec.~\ref{sec4}, for an initial stage of the process, we calculate a scale for which the generation of large scale magnetic fields is the most intense.

In Ref.~\cite{Dvo25}, by making numerical simulations, we received that lower lateral harmonics of the magnetic field have greater amplitudes compared to higher ones. We put forward an assumption that it was the indication to some instability in axionic MHD present in Eq.~\eqref{meaN}. Now, we can provide a possible explanation of this feature. When the mirror symmetry of the axion soliton is broken, the $\gamma$-term in the
axionic mean field equation interacts with principal harmonics. As the result, the axion soliton acquires the magnetic energy and becomes helical.
We have constructed a simplest statistical ensemble $\Omega$, which is defined by the projection of all harmonics onto principal harmonics. 

The instability of the axion soliton
is related with a growth of lateral harmonics. In our statistic ensemble, this process in not
allowed, because this process is related with the $\alpha$-term in the mean field equation, which is projected in $\Omega$ into zero. By this effect, we adjust the power of the energy to the $\alpha$-term rather than to the $\gamma$-term, i.e. we assume that the scalar term in the axion energy interacts with lateral harmonics more intensively. 
The power of the $\gamma$-term at the initial stage of the process tends to zero more rapidly when the wave number incises. At the initial stage of the process, when the axion soliton takes magnetic energy, the helicity generation for principal harmonics is weak. 

Before the phase transition, when the energy of the system is concentrated in the axion component, the ensemble of the main magnetic harmonics of the Hopf system is represented by the high frequency harmonics with a long spectral interval, the $L^2$-norm of the representative is small. It is the assumption that the magnetic helicity density with the main term $(\A,\B)$ disappears in the domain, while the quadratic helicity density with the main term $(\A,\B)^2$ is not assumed to be small. Let us assume   that the quadratic helicity flux caused by the mean field Eq.~\eqref{meaN} is uniformly distributed over the spectrum of the main harmonics of the system.
 	Using this assumption, we are able to prove that the magnetic spectral index is determined by the quadratic magnetic helicity spectral index and coincides with the Iroshnikov-Kraichnan spectrum index. This statement results from Ref.~\cite{A1}. However, the formal proof is still absent.
 		
 	An axion bunch is characterized by the total energy and the ratio of the energies of the axion and magnetic components, which is the value of the densities of the first two moments of magnetic helicity. It is likely that such a bunch is also characterized by higher moments of the magnetic helicity. 
 	The process of the magnetic energy release from a twisting soliton has not been fully studied in the present work. We assume that the twisted soliton interacts with matter and obeys the Kolmogorov law, which determines the third term in Eq.~\eqref{meaN}. Such an interaction is described both through the norm of the inhomogeneous axion potential $\varphi$ and through the norm
 	of its gradient $\nabla \varphi$. We recall that the axion norm is measured in a non-classical way as the norm in the corresponding Sobolev space. When deriving the formula for the wave number, we took into account only the interaction associated with the potential $\varphi$ (weaker), and we did not take into account the interaction with the norm $\nabla \varphi$, which can also occur under the linear assumption at a later stage.

The spectrum of the $M$-invariant has the density, which coincides with the flux index of the Kolmogorov spectrum. To describe this spectrum it is necessary to calculate the highest moments of magnetic line coupling. However, this study is beyond the scope of the work. Numerical simulations, carried out in Ref.~\cite{Dvo25}, showed that the phase transition is accompanied by an increase in lateral harmonics, which only at a subsequent stage lead to an increase in the density of the magnetic helicity. More precisely, the process of phase transition in this domain should be determined by the numerical solution of Eq.~\eqref{meaN} together with the solution of the Klein-Gordon equation for the axion. Up to now, this is result is obtained only approximately.
 	
To definitively answer to the question how the wave number, calculated for the magnetic field in Sec.~\ref{sec4}, is related to the characteristics of the axion bunch, one has to carry out numerical simulations and write down an equation for the magnetic helicity flux that depends on the higher moments. We were able to do this only under the assumption of a constant magnetic helicity flux density after the phase transition, which is apparently a correct simplifying assumption. However, it does not take into account complex physics concepts behind this problem.

While making numerical simulations in Refs.~\cite{Dvo24,Dvo25}, we decomposed both axionic and magnetic fields over the same set of the latitude harmonics. Moreover, axionic and magnetic fields were supposed to evolve within the same spherically symmetric region. Thus, in those particular studies, the length scales of these fields should coincide. Thus, the wave number corresponding to the maximum of the spectrum, obtained in Sec.~\ref{sec4}, should be related to the reciprocal size of the axionic clump. However, as mentioned earlier, a careful proof of this statement has not been made yet.

\appendix

\section{The generalized Arnold inequality in the case of random distributions of magnetic fields}\label{sec5}

The inequality in Eq.~(\ref{BB}) shows that the quadratic magnetic helisity $\chi^{(2)}_{\B}$, which is invariant in the ideal MHD, gives a lower bounds  for $U^{(4)}_{\B}$. To prove this fact, we collect inequalities in Eqs.~(\ref{chi(2)}) and~(\ref{Uchi(2)})
as following:
\begin{eqnarray}\label{B(4)}
	U^{(4)}_{\B} \ge C^2 \chi^{(2)}_{\B}, 
\end{eqnarray}
where $C$ is a constant described in Theorem \ref{Arnold}.
The inequality in Eq.~(\ref{Uchi(2)}) is not optimal at least in the case when a random magnetic field $\B$ is distributed as a quasi-periodic vector-function with a prescribed power for magnetic energy distribution: $\B^2(k) \sim k^{\aleph}$, where
$\aleph \le -1$ is a parameter, which is called a power of the distribution and wave number $k > \delta_0$, where $\delta_0$ is a lower bound of the magnetic spectra. The question arises whether the constant in Eq.~(\ref{B(4)}) can be greater.

With the assumption above, the scalar function $(\A,\B)^2$ is distributed over the $k$-line, $k > \delta_0$, as 
$(\A,\B)^2 \sim k^{2\aleph - 2}$ and $(\B,\B)^2$ is distributed over the $k$-line as
$(\B,\B)^2 \sim k^{2\aleph}$. The distribution of the quadratic magnetic helicity $\chi^{(2)}_{\B}$ is calculated as the square of the
distribution of the series: 
\begin{equation}\label{chi}
	\chi(x) = \lim_{a \to +\infty}\sum_{i=0}^{\infty} \frac{a^i}{i!} [ (\nabla_{\B})^{2i} (\A(x),\B(x))].
\end{equation}
Assume that the main term in Eq.~(\ref{chi}) is distributed with zero average. In this case, $\chi_{\B}=0$.
The total distribution in Eq.~(\ref{chi}) is proportional to the (central) distribution of the main term $i=0$ with the coefficient
$\frac{\sqrt{\pi}}{2}$. This proves that the following distributions coincide:  
$\chi^{(2)}_{\B} = \frac{\pi}{4}(\A,\B)^2$. 
Therefore we get the following inequality:
\begin{eqnarray}\label{BB(4)}
	U^{(4)}_{\B} \ge \frac{4 C^2}{\pi} \chi^{(2)}_{\B}, 
\end{eqnarray}
which is stronger then Eq.~(\ref{B(4)}).

\section{The Biot-Savart's law: A particular case of Maxwell Theorem on multipoles}\label{sec6}

Define:
$$\A(\x_2,\x_1)= \frac{1}{4\pi} \frac{\B(\x_2) \times (\x_1-\x_2)}{\vert\vert \x_1 - \x_2 \vert\vert^3}. $$ 
The vector-function
$\A(\x_2,\x_1)$ with a singularity in  $\x_2$ is called the Biot-Savart vector-potential.

\begin{lemma}\label{BioSavart}
	Let $\B=\B(\x_1)$ be the magnetic field in $\Omega \subset \R^3$, $\x_1 \in \Omega$, which is tangent to the boundary  $\partial \Omega$. Let us define the vector-potential by the formula: 
	$\A(\x_1) = \int_{\Omega} \A(\x_2;\x_1) d\x_2$. Then, $\A(\x_1) \to 0$ at $\x_1 \to +\infty$. 
	Moreover, the following equality is satisfied:
	$$ \rot (\A(\x_1)) = \B(\x_1), \quad \x_1 \in \Omega. $$
\end{lemma}

\subsubsection*{Proof of Lemma \ref{BioSavart}}
Define $\W(\x_2;\x_1)=\frac{-\B(\x_1)}{4 \pi \vert \x_1 - \x_2 \vert}$, 
$\W(\x_1) = \int_{\Omega} \W(\x_1;\x_2) d \x_2$. We have:
$\rot \rot(\W(\x_1)) = -\Delta (\W(\x_1)) + \grad \div(\W(\x_1)) =\B(\x_1)$, because
$\div (\W(\x_1)) = 0$, this is a corollary of $\div(\B(\x_1))=0$. One has that
$\W(\x_2;\x_1)$ is the fundamental function of the Laplace operator, which is
associated with the vector
$\delta$-function with the source at $\B(\x_2)$. Recall that the vorticity operator $\rot$
is taken over $\x_1$. On the other hand,   $\rot \W(\x_2;\x_1) = \B(\x_2) \times \grad( \frac{1}{4 \pi \vert \x_1 - \x_2 \vert}) =$ $\frac{1}{4 \pi}\B_2 \times \frac{\x_1 - \x_2}{\vert \x_1 - \x_2 \vert^3} = \A(\x_1,\x_2)$. \qed    
\[  \]

The next example shows that the vorticity operator is not commuting with the mean operator
of an elementary dipole magnetic source.
The reason is the following: for the elementary dipole source exists a singularity.
For the mean value of the vorticity vector the Weierstrass condition of uniformly
convergence is not satisfied. This example is present in Ref.~\cite{A3}. The Maxwell Multipole Theorem in Ref.~\cite{Arn2} gives a lot of analogous examples, which are not investigated and explicitly formulated. 

Let  $\Omega$ be an infinitesimal thin magnetic tube (a closed magnetic line) of 
$\B$, 
$\A(\x), \quad \x \in \Omega$ be the magnetic vector-potential, which is defined by the Biot-Savart formula:
\begin{equation} 
\A(\x) = \int_{\Omega} \A(\x,\x_1) d\x_1. \label{int}
\end{equation}
 To analyze Eq.~\eqref{int} we mention that the integration over the parameter
$\x_1$ on the magnetic line (the avarage operator) and the vorticity operator $\rot$ over the variable $\x \in \R^3$ 
are not commuted. The following expression is satisfied:
\begin{eqnarray}\label{2/3}
	\rot (\A(\x)) = \frac{2}{3} \int_{\Omega} \rot (\A(\x,\x_1)) d \x_1, 
\end{eqnarray}
where the vector-function in the integral at the right hand side of the equation is
limit in $\x$ of $\x_1$-family of vector-functions
$\rot (\A(\x,\x_1))$ in terms of principal value of  improper integrals, as is proved in  Ref.~\cite{A3}.
Consider an $\x_1$--family of vector-functions 
$\rot_{\x}(\varphi(\x) \A(\x,\x_1))$ 
and use the Gauss-Ostrogradskii formula for this family. 
As the result, the integrals of the functions 
$\div(\rot (\A(\x,\x_1))\varphi(\x))$ and $\div(\rot(\A(\x,\x_1))\times \grad(\varphi(\x)))$  
are related by residues at the critical points $\x_1$. It proves the validity of Eq.~(\ref{2/3}).


\begin{thebibliography}{50}

	
	\bibitem{Tis07}
	P.~Tisserand et al. (EROS-2 Collaboration),
	Limits on the Macho Content of the Galactic Halo from the EROS-2 Survey of the Magellanic Clouds,
	Astron. Astrophys. \textbf{469}, 387--404 (2007)
	[astro-ph/0607207].
	
	\bibitem{PecQui77}
	R.~D.~Peccei and H.~R.~Quinn,
	CP Conservation in the Presence of Pseudoparticles,
	Phys. Rev. Lett. \textbf{38}, 1440--1443 (1977).
	
	\bibitem{Wei78}
	S.~Weinberg,
	A New Light Boson?,
	Phys. Rev. Lett. \textbf{40}, 223--226 (1978).
	
	\bibitem{Wil78}
	F.~Wilczek,
	Problem of Strong P and T Invariance in the Presence of Instantons,
	Phys. Rev. Lett. \textbf{40}, 279--282 (1978).
	
	\bibitem{Abe20}
	C.~Abel et al.,
	Measurement of the Permanent Electric Dipole Moment of the Neutron,
	Phys. Rev. Lett. \textbf{124}, 081803 (2020)
	[arXiv:2001.11966].
	
	\bibitem{DinFis83}
	M.~Dine and W.~Fischler
	The not-so-harmless axion
	Phys. Lett. B \textbf{120} 137--141 (1983).
	
	\bibitem{KolTka93}
	E.~W.~Kolb and I.~I.~Tkachev,
	Axion miniclusters and Bose stars,
	Phys. Rev. Lett. \textbf{71}, 3051--3054 (1993)
	[hep-ph/9303313].
	
	\bibitem{KolTka94}
	E.~W.~Kolb and I.~I.~Tkachev,
	Non-Linear Axion Dynamics and Formation of Cosmological
	Pseudo-Solitons,
	Phys. Rev. D \textbf{49}, 5040--5051 (1994)
	[astro-ph/9311037].
	
	\bibitem{BraZha19}
	E.~Braaten and H.~Zhang,
	Colloquium: The physics of axion stars,
	Rev. Mod. Phys. \textbf{91}, 041002 (2019).
	
	\bibitem{Vis21}
	L.~Visinelli,
	Boson Stars and Oscillatons: A Review,
	Int. J. Mod. Phys. D \textbf{30}, 2130006 (2021)
	[arXiv:2109.05481].
	
	\bibitem{KimCar10}
	J.~E.~Kim and G.~Carosi,
	Axions and the strong CP problem,
	Rev. Mod. Phys. \textbf{82}, 557--601 (2010)
	[arxiv:0807.3125].
	
	\bibitem{LonVac15}
	A.~Long and T.~Vachaspati,
	Implications of a primordial magnetic field for magnetic monopoles, 
	axions, and Dirac neutrinos,
	Phys. Rev. D \textbf{91}, 103522 (2015)
	[arXiv:1504.03319].
	
	\bibitem{DvoSem20}
	M.~Dvornikov and V.~B.~Semikoz,
	Evolution of axions in the presence of primordial magnetic fields,
	Phys. Rev. D \textbf{102}, 123526 (2020)
	[arxiv:2011.12712].
	
	\bibitem{Dvo22}
	M.~Dvornikov,
	Interaction of inhomogeneous axions with magnetic fields
	in the early universe,
	Phys. Lett. B \textbf{829}, 137039 (2022)
	[arxiv:2201.10586].
	
	\bibitem{Anz23}
	F.~Anzuini, J.~A.~Pons, A.~G\'omez-Ba\~n\'on, P.~D.~Lasky, F.~Bianchini, and A.~Melatos,
	Magnetic Dynamo Caused by Axions in Neutron Stars,
	Phys. Rev. Lett. \textbf{130}, 071001 (2023)
	[arxiv:2211.10863].
	
	\bibitem{Bey23}
	K.~A.~Beyer, G.~Marocco, C.~Danson, R.~Bingham, and G.~Gregori, 
	Parametric co-linear axion photon instability,
	Phys. Lett. B \textbf{839}, 137759 (2023)
	[arxiv:2108.01489].
	
	\bibitem{D-A}
	M.~S.~Dvornikov and P.~M.~Akhmet'ev,
	Magnetic field evolution in spatially inhomogeneous axion structures, 
	Theor. Math. Phys. \textbf{218}, 515--529 (2024).
	
	\bibitem{AkhDvo24}
	P.~Akhmetiev and M.~Dvornikov,
	Magnetic fields in inhomogeneous axion stars,
	Int. J. Mod. Phys. D \textbf{33}, 2450001 (2024)
	[arxiv:2303.09254].
	
	\bibitem{Dvo24}
	M.~Dvornikov,
	Thin layer axion dynamo,
	Eur. Phys. J. C \textbf{84}, 892 (2024)
	[arxiv:2401.03185].
	
	\bibitem{Dvo25}
	M.~Dvornikov,
	Low mode approximation in the axion magnetohydrodynamics,
	Int. J. Mod. Phys. D 
	(2025),
	doi:~10.1142/S0218271825410019
	[arxiv:2502.20839].

\bibitem{Kli15}
  J.~A.~Klimchuk,
  Key Aspects of Coronal Heating,
  Phil. Trans. R. Soc. A: Math. Phys. Eng. Sci. \textbf{373}, 20140256 (2015)
  [arXiv:1410.5660].
	
\bibitem{DilZio03}
  L.~DiLella and K.~Zioutas,
  Observational evidence for gravitationally trapped massive axion(-like) particles,
  Astropart. Phys. \textbf{19}, 145--170 (2003)
  [astro-ph/0207073].

\bibitem{Zhi17}
  A.~R.~Zhitnitsky,
  Solar Extreme UV radiation and quark nugget dark matter model,
  J. Cosmol. Astropart. Phys. \textbf{10}, 050 (2017)
  [arXiv:1707.03400].
	
	\bibitem{D-A1}
	M. S. Dvornikov and P. M. Akhmetiev,  {\it Evolution of a Large Scale Chern-Simons Magnetic Field
		in Expanding Space with a Negative Scalar Curvature Parameter}, Physics of Particles and Nuclei, 2025, Vol. 56, No. 2, pp. 269--274.
	
	\bibitem{Arn}
	V.I.Arnold, {\it The asymptotic Hopf invariant and its applications}, Proc. Summer School in Diff. Equations, 1973 (1974), Erevan (in Russian); Sel. Math. Sov. 5, 327-345 (1986).
	
	\bibitem{A1}
	P.M.~Akhmet'ev, {\it Quadratic Helicities and the Energy of Magnetic Fields}
	Proceedings of the Steklov Institute of Mathematics, , Vol. 278, pp. 10-21 (2012).
	
	\bibitem{D}
	P.~Dehornoy, {\it Asymptotic invariants of 3-dimensional vector fields}, Winter Braids Lecture Notes, 2 (2016), 1
	
	\bibitem{S}
	D.D.~Sokoloff, {\it Mirror Asymmetry and Helicity Invariants in Astrophysical Dynamos}, Geomagn. Aeron., 59:7 (2019), 799-805
	
	\bibitem{A2}
		P. M. Akhmet'ev, {\it Magnetic helicity flux for mean magnetic field equations}, TMF, 204:1 (2020), 130--141; Theoret. and Math. Phys., 204:1 (2020), 947--956.

	
	\bibitem{A3}
		P. M. Akhmet'ev, {\it A higher-order analog of the helicity number for a pair of divergent-free vector fields}, Geometry and topology. Part 6, Zap. Nauchn. Sem. POMI, 279, POMI, St. Petersburg, 2001, 15--23; J. Math. Sci. (N. Y.), 119:1 (2004), 5--9

	\bibitem{A-K-S}	
	P. M. Akhmet'ev, E. A. Kudryavtseva, and A. Yu. Smirnov, {\it A generalization of the Arnol'd inequality in MHD}, Magnetohydrodynamics, 52:1 (2016), 5-14
	
	
	\bibitem{Arn2}
	 V.~Arnold,
	{\it Topological content of the Maxwell theorem on multipole representation of spherical functions},
	Topol. Methods Nonlinear Anal. 7, no. 2, 205-217 (1996).
	
	\bibitem{M-F}
	A. S. Mishchenko, A. T. Fomenko, Euler equations on finite-dimensional Lie groups, Math. USSR-Izv., 12:2 (1978), 371--389

	
	\bibitem{Mo}
	H. K. Moffatt, {\it Magnetic Field Generation in Electrically Conducting Fluids}, Cambridge Univ. Press, Cambridge (1978). 
	
	\bibitem{A-Kh}
	V.~I.~Arnold and B.~A.~Khesin, {\it Topological Methods in Hydrodynamics}, 
Applied Mathematical Sciences vol 125 (1998)(2013) (New York:
Springer-Verlag).
	
	\bibitem{Z-R-S}
	Ya.B. Zeldovich, A.A. Ruzmaikin, and D.D. Sokoloff, {\it Magnetic Fields in Astrophysics}, Gordon and Breach, New York, 1983.
	
	
	
\end{thebibliography}
\end{document}